\documentclass{emulateapj}
\usepackage{apjfonts}

\newcommand{\be}{\begin{equation}}
\newcommand{\ee}{\end{equation}}

\shorttitle{High $\ell$ $TT$ results from QUaD}
\shortauthors{QUaD collaboration}

\begin{document}

\slugcomment{Submitted to ApJL}

\title{Small Angular Scale Measurements of the CMB Temperature Power Spectrum from QUaD}

\author{
  QUaD collaboration
  --
  R.\,B.\,Friedman\altaffilmark{1},
  P.\,Ade\altaffilmark{2},
  J.\,Bock\altaffilmark{3,4},
  M.\,Bowden\altaffilmark{2,5},
  M.\,L.\,Brown\altaffilmark{6},
  G.\,Cahill\altaffilmark{7},
  P.\,G.\,Castro\altaffilmark{8,9},
  S.\,Church\altaffilmark{5},
  T.\,Culverhouse\altaffilmark{1},
  K.\,Ganga\altaffilmark{10},
  W.\,K.\,Gear\altaffilmark{2},
  S.\,Gupta\altaffilmark{2},
  J.\,Hinderks\altaffilmark{5,11},
  J.\,Kovac\altaffilmark{4},
  A.\,E.\,Lange\altaffilmark{4},
  E.\,Leitch\altaffilmark{3,4},
  S.\,J.\,Melhuish\altaffilmark{12},
  Y.\,Memari\altaffilmark{8},
  J.\,A.\,Murphy\altaffilmark{7},
  A.\,Orlando\altaffilmark{2,4}
  C.\,O'\,Sullivan\altaffilmark{7},
  L.\,Piccirillo\altaffilmark{12},
  C.\,Pryke\altaffilmark{1},
  N.\,Rajguru\altaffilmark{2,13},
  B.\,Rusholme\altaffilmark{5,14},
  R.\,Schwarz\altaffilmark{1},
  A.\,N.\,Taylor\altaffilmark{8},
  K.\,L.\,Thompson\altaffilmark{5},
  A.\,H.\,Turner\altaffilmark{2},
  E.\,Y.\,S.\,Wu\altaffilmark{5}
  and
  M.\,Zemcov\altaffilmark{2,3,4}
}

\altaffiltext{1}{Kavli Institute for Cosmological Physics,
  Department of Astronomy \& Astrophysics, Enrico Fermi Institute, University of Chicago,
  5640 South Ellis Avenue, Chicago, IL 60637, USA.}
\altaffiltext{2}{School of Physics and Astronomy, Cardiff University,
  Queen's Buildings, The Parade, Cardiff CF24 3AA, UK.}
\altaffiltext{3}{Jet Propulsion Laboratory, 4800 Oak Grove Dr.,
  Pasadena, CA 91109, USA.}
\altaffiltext{4}{California Institute of Technology, Pasadena, CA
  91125, USA.}
\altaffiltext{5}{Kavli Institute for Particle Astrophysics and
Cosmology and Department of Physics, Stanford University,
382 Via Pueblo Mall, Stanford, CA 94305, USA.}
\altaffiltext{6}{Cavendish Laboratory,
  University of Cambridge, J.J. Thomson Avenue, Cambridge CB3 OHE, UK.}
\altaffiltext{7}{Department of Experimental Physics,
  National University of Ireland Maynooth, Maynooth, Co. Kildare,
  Ireland.}
\altaffiltext{8}{Institute for Astronomy, University of Edinburgh,
  Royal Observatory, Blackford Hill, Edinburgh EH9 3HJ, UK.}
\altaffiltext{9}{{\em Current address}: CENTRA, Departamento de F\'{\i}sica,
  Edif\'{\i}cio Ci\^{e}ncia, Piso 4,
  Instituto Superior T\'ecnico - IST, Universidade T\'ecnica de Lisboa,
  Av. Rovisco Pais 1, 1049-001 Lisboa, Portugal.}
\altaffiltext{10}{APC/Universit\'e Paris 7 - Denis
  Diderot/CNRS, B\^atiment Condorcet, 10, rue Alice Domon et L\'eonie
  Duquet, 75205 Paris Cedex 13, France}
\altaffiltext{11}{{\em Current address}: NASA Goddard Space Flight
  Center, 8800 Greenbelt Road, Greenbelt, Maryland 20771, USA.}
\altaffiltext{12}{{\em Current address}: School of Physics and
  Astronomy, University of
  Manchester, Manchester M13 9PL, UK.}
\altaffiltext{13}{{\em Current address}: Department of Physics and Astronomy, University
  College London, Gower Street, London WC1E 6BT, UK.}
\altaffiltext{14}{{\em Current address}:
  Infrared Processing and Analysis Center,
  California Institute of Technology, Pasadena, CA 91125, USA.}

\begin{abstract}
We present measurements of the cosmic microwave 
background (CMB) radiation temperature anisotropy in the multipole
range $2000<\ell<3000$ from the QUaD 
telescope's second and third observing seasons.  
After masking the brightest point sources our results are 
consistent with the primary $\Lambda$CDM expectation alone.
We estimate the contribution of residual (un-masked) radio point
sources using a model calibrated to our own bright source observations,
and a full simulation of the source finding and masking procedure.
Including this contribution slightly improves the $\chi^2$.
We also fit a standard SZ template to the bandpowers and see no strong 
evidence of an SZ contribution, which is as expected for 
$\sigma_8 \approx 0.8$.
\end{abstract}
\keywords{cosmology: cosmic microwave background,
cosmology: observations}

\section{Introduction}

Observations of the cosmic microwave background (CMB) anisotropy at
angular scales of several arcminutes or larger ($\ell<2000$) have been used 
to constrain parameters of the $\Lambda$CDM cosmological model to high 
precision \citep{castro09,dunkley08}.
At these larger angular scales, the anisotropic power is dominated by the 
primary CMB from the surface of last scattering.
At smaller angular scales ($\ell>2000$) the primary anisotropy is
exponentially suppressed by diffusion in the primordial plasma 
and the structure becomes dominated by foreground emission 
and secondary anisotropy generated by intervening large scale structure.
Measuring the secondary anisotropy introduced by the thermal 
Sunyaev-Zel'dovich effect (SZE) has been of particular interest.
The magnitude of the SZE power is a sensitive and independent probe 
of the amplitude of density perturbations, 
scaling~as~$\sigma_8^7$~\citep{komatsu02}.

Previous measurements of the small angular scale CMB anisotropy at 
30~GHz by CBI \citep{readhead04b} and BIMA \citep{dawson06} claimed a 
significant excess over the $\Lambda$CDM expectation at multipoles of 
$\ell>2000$.
The ACBAR experiment \citep{reichardt08} subsequently reported a 
$\sim1$-$\sigma$ excess at 150~GHz at similar scales.
Attributing this excess power to the SZE alone implies $\sigma_8\approx1$. 
This value is in conflict with the WMAP 5-year results \citep{dunkley08} and 
recent X-ray measurements of the cluster mass function \citep{vikhlinin08}, 
which both yield values of $\sigma_8\approx0.8$.
For the latter value of $\sigma_8$, the SZE power at 30~GHz is expected to 
be comparable to the primary CMB at multipoles of $\ell\approx2500$ but at 
100 and 150~GHz will be subdominant at multipoles of $\ell<3000$.
The results presented in this work cover a multipole range of 
$2000<\ell<3000$ and are the highest sensitivity to date at these scales.

The QUaD telescope is a millimeter-wavelength bolometric polarimeter 
located at the South Pole. 
QUaD operated during the austral winters of 2005 to 2007.
Details of the QUaD instrument, calibrations and performance can 
be found in \cite{hinderks08} and \cite{osullivan08}.
In this paper we present high-$\ell$ {\it TT} spectra only.
Details of the observations, data quality, low level processing, map-making 
and power spectrum estimation plus the full polarization analysis for 
$\ell<2000$ can be found in \cite{pryke09} and \cite{brown09}
--- this paper follows the analysis 
methods described there except where noted.

\section{Analysis}
\label{sec:analysis}

In order to reduce the bandpower uncertainty at high-$\ell$
we have adopted an optimal signal to noise 
Fourier plane weighting step in the power spectrum estimation.
As can be seen in Figure~7 of \cite{pryke09}, the distribution 
of noise power in the two-dimensional Fourier plane is highly non-uniform
--- the atmospheric noise forms a concentrated band around 
the y-axis.
Fourier plane weights are calculated as
\begin{equation}
F=\frac{S_{\mathrm{CMB}}^2}{(S_{\mathrm{CMB}}+N)^2},
\end{equation}
where $S_{\mathrm{CMB}}$ and $N$ are the ensemble averages of 
the signal and noise simulation two dimensional auto power spectra.
Since the CMB signal is expected to be uniformly distributed in azimuth 
angle, downweighting localized regions of high noise in the Fourier plane 
will not  bias the results so long as the weighting is independent
of the actual data values.
This weighting has a dramatic effect on the bandpower 
uncertainty at high-$\ell$ --- for 150~GHz the error is suppressed by 
as much as an order of magnitude in the range of $2500<\ell<3000$.

\begin{figure}[t!]
\begin{center}
\resizebox{\columnwidth}{!}{\includegraphics{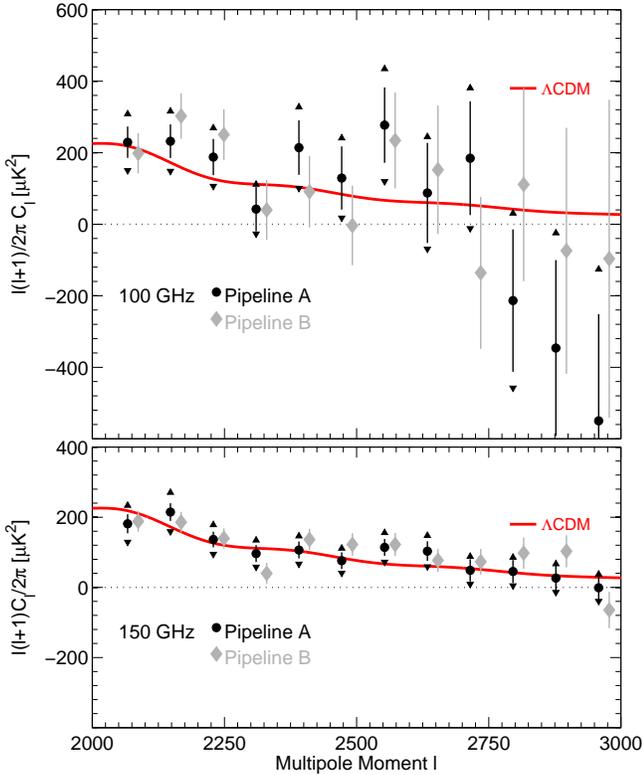}}
\end{center}
\caption{
{\it TT} bandpower values at 100~GHz ({\it top}) and 150~GHz 
({\it bottom}) versus $\Lambda$CDM.
We show results using the same data from two independent pipelines
(see text); the points are offset for clarity. 
The error bars include $\Lambda$CDM sample variance and noise only.
The triangles indicate the coherent shift in the errorbar end points
which result when the beam and absolute calibration are simultaneously
pushed up/down by 1-$\sigma$.
}
\label{fig:simerrb}
\end{figure}

\begin{figure}
\begin{center}
\resizebox{\columnwidth}{!}{\includegraphics{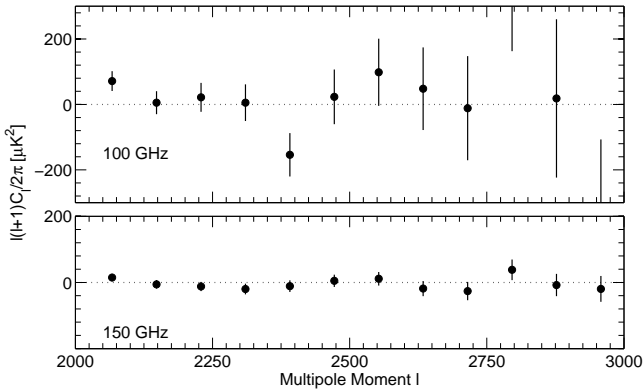}}
\end{center}
\caption{
Jackknife {\it TT} bandpower values for both the 100~GHz ({\it top}) 
and 150~GHz ({\it bottom}) deck-jackknife maps.
These spectra are consistent with null and indicate that the
signal seen in Figure~\ref{fig:simerrb} originates on the sky.
}
\label{fig:dkjack}
\end{figure}

In this paper we also use two enhancements over the \cite{pryke09}
analysis introduced in \cite{brown09}.
Firstly we replace the field differencing operation used to remove 
ground contamination with a template removal technique
that doubles our effective sky area, further reducing the 
bandpower uncertainties by a factor of $\sim\sqrt2$.
Secondly the beam model used in this paper has been
updated to include sidelobes measured at the 
$<-20$dB level and predicted by the physical optics simulations
described in \cite{osullivan08}.
The absolute calibration uncertainty is now 7\% in power.
The cosmological model assumed in our simulations and analysis 
is the WMAP 5-year model given in column two of Table~2 in~
\cite{dunkley08} with zero SZE signal, hereafter referred to as 
$\Lambda$CDM.
The input simulation spectra are generated using CAMB \citep{lewis99} with 
lensing turned on.

\section{Results}
\label{sec:results}

In Figure~\ref{fig:simerrb} we present our basic result, the {\it TT} band 
power values at 100 and 150~GHz extending to $\ell$~=~3000.
Pipeline A is the pipeline used in \cite{pryke09},
while Pipeline B is an alternate curved sky analysis (see \citealt{ade07}
and \citealt{brown09}).
Both pipelines are based broadly on the MASTER analysis technique 
\citep{hivon02}.
The bandpower uncertainties are calculated from the spread in signal 
plus noise simulation bandpowers assuming the
input $\Lambda$CDM theory spectrum shown.

As discussed in detail in \cite{pryke09}, jackknife maps made from 
differencing independent data sets covering the same sky are a powerful
test for systematic contamination.
In Figure~\ref{fig:dkjack} we present bandpower values for the 
deck-jackknife --- likely our most stringent test.
These jackknife spectra are consistent with null.

In the multipole range considered for this analysis, the sky power has 
been suppressed by almost an order of magnitude through beam 
convolution.
Thus a small mis-estimate of the beam would result in a large, 
multipole--dependent systematic shift in the bandpower values.
An under(over)-estimate of the beam suppression would result in an 
under(over)-estimate of our bandpowers.
The effect of the systematic uncertainties on our beam model and 
calibration is illustrated in Figure~\ref{fig:simerrb} where we show
the result of pushing both up/down simultaneously by 1-$\sigma$.
While systematic uncertainty is significant at both frequencies,
it is not sufficient to qualitatively change the results at 150~GHz.

\begin{figure}
\begin{center}
\resizebox{\columnwidth}{!}{\includegraphics{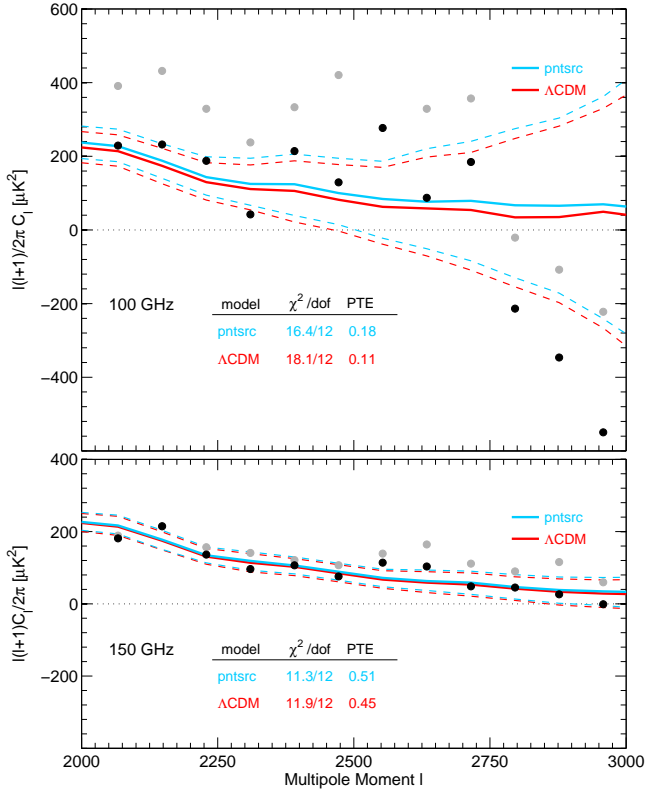}}
\end{center}
\caption{
{\it TT} bandpower values at 100~GHz ({\it top}) and 150~GHz
({\it bottom}).
This Figure shows the same information as Figure~\ref{fig:simerrb}
but instead of using error bars we plot the spread in the signal plus
noise simulations versus bare points from the data.
This is in a sense the fundamental result of the analysis --- is
the data consistent with being a realization of the simulation model?
The lines show the 16, 50 and 84\% points of the simulation distribution
(corresponding to -1, 0, +1 $\sigma$ for a Gaussian distribution).
Simulation distributions are shown for $\Lambda$CDM alone and for
$\Lambda$CDM plus a residual radio source foreground ({\it pntsrc}).
The $\chi^2$ values are calculated for the data versus the simulation 
model (see text).
We also plot bandpowers calculated without masking sources in the
maps as {\it light points}.
}
\label{fig:simband}
\end{figure}

Though it is customary to present the bandpower results as in 
Figure~\ref{fig:simerrb}, it is arguably more natural to consider them as we 
do in Figure~\ref{fig:simband}.
We calculate $\chi^2$ between the data and the simulation distributions
using the spread in the signal plus noise realizations to construct a 
bandpower covariance matrix \citep[see][]{pryke09,brown09}.
The resulting $\chi^2$ values calculated over the $\ell$ bins presented are 
shown in Figure~\ref{fig:simband}.

\setcounter{footnote}{0}
The bandpowers presented in Figures~\ref{fig:simerrb}~\&~\ref{fig:simband}
were calculated after masking bright point sources in the maps as 
described in Section~\ref{sec:psims}.
We detect 7 point sources at $>5\sigma$ ($\sim50$mJ) in both the 100 and 
150~GHz maps; 
all of these were matched with a low-probability of chance-association
to PMN \citep{pmn} or SUMSS \citep{sumss} radio sources
using NED\footnote{http://nedwww.ipac.caltech.edu/}. 
The effect of masking them can be seen in Figure~\ref{fig:simband} 
as the difference between the light and dark points.
Since radio source populations typically follow a power law
distribution these sources are only the sparse high flux end of an 
exponentially more numerous low flux population.
It is therefore necessary to estimate the residual power contribution 
from the unmasked radio source population.

\section{Point Source Simulations}
\label{sec:psims}

The effect of residual point source contamination in our
spectra will manifest as both an increase in the total power at a given 
$\ell$ and an increase in the bandpower fluctuation.  
Though it would be straightforward to estimate the mean power 
contribution from a given point source model and flux cut analytically 
the subtle effects of source identification and masking in our pipeline 
would be difficult to account for accurately.  
Moreover, the fluctuations are potentially non-Gaussian.
Instead, we explicitly simulate the source population in our maps.

\begin{figure}
\begin{center}
\resizebox{\columnwidth}{!}{\includegraphics{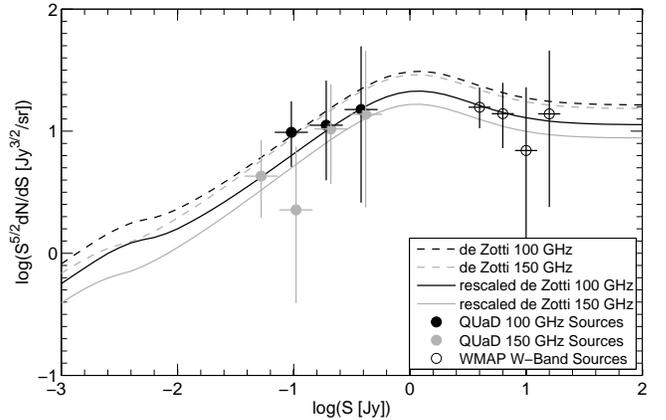}}
\end{center}
\caption{
The simulation input point source model ({\it solid lines}) is derived 
by applying a simple rescale factor to the de Zotti radio point source model 
({\it dashed lines}) \citep{dezotti05} at each frequency.  
The rescale factor is determined from fits to $dN/dS$
points derived from sources detected in our maps ({\it closed circles}).  
At 100~GHz, the re-scaling also brings the de Zotti model into better 
agreement with WMAP W-band $dN/dS$ points ({\it open circles}).
}
\label{fig:dnds}
\end{figure}

At present, the statistical properties of the sub-Jansky radio source 
population are not well known at 100 and 150~GHz.
The most useful published data is the W-band (94GHz) WMAP point source 
catalog \citep{wright08}.
However, this catalog is only complete at fluxes at or above several 
Jansky whereas even the brightest sources in the QUaD maps are 
sub-Jansky.
Therefore to predict the radio source distribution below our detection 
threshold we use the \cite{dezotti05} extragalactic radio source model.
Although this is a carefully constructed model using detailed 
astrophysics, it's calibration at lower frequencies makes it a distant 
extrapolation at 150~GHz.
We rescale the model by 0.7 and 0.6 at 100 and 150~GHz respectively 
to match the number of sources observed in the QUaD maps
(see Figure~\ref{fig:dnds}).
Doing so also brings the model into better agreement with crude $dN/dS$ 
points determined from the WMAP catalog.
We have ignored the possible contribution from IR (dusty) sources.

For point sources, unlike the CMB, we do not simulate the sky signal at the 
timestream level.
Instead we directly inject source populations into existing signal
plus noise maps as beam sized blips.  
The fluxes are drawn from the source 
population model and the positions are uniformly distributed across
the map area, i.e. no clustering is assumed.

While sources of low to moderate flux are abundant those of high flux are 
rare.
Leaving the brightest sources in the maps leads to huge contamination
as we see comparing the black and grey points in Figure~\ref{fig:simband}.
Thus we must mask sources which are detected at high significance.
Given a map containing three components --- CMB, noise and point 
sources --- we need an automated and unbiased method for identifying the 
point sources to mask.

To separate out the point sources in our maps we adopt a two-dimensional 
Fourier space optimal filter (Weiner filter) given by
\begin{equation}
\label{eq:weiner}
W=\frac{S_{pnt}}{S_{pnt}+S_{\mathrm{CMB}}+N},
\end{equation}
where $S_{pnt}$ is the point source signal.
We fix $S_{pnt}$ to be a beam suppressed white-noise power spectrum 
--- representing a uniform distribution of point sources --- 
with an amplitude at $\ell\sim2500$ roughly equal to our bandpower 
value at that multipole.
The resulting filters are broadly azimuthally symmetric in form.
They are non-zero in the multipole range where we are most sensitive to 
point source power 
--- i.e. zero at the lowest multipoles where CMB signal is dominant 
then rising to a peak near $\ell\sim2000$ for 100~GHz and $\ell\sim2500$ 
for 150~GHz before decaying back to zero at the highest multipoles where 
instrumental noise is dominant.

The filtered map is then given by
\begin{equation}
\label{eq:map}
m'=\textrm{FT}\bigg(W \cdot \textrm{FT}\bigg(\frac{m}{v_{pix}}\bigg)\bigg)\cdot v_{pix},
\end{equation}
where FT is the Fourier transform operation, $m$ is the original
map, $W$ is the Weiner filter defined in equation~\ref{eq:weiner} 
and $v_{pix}$ is the pixel variance map --- used here
to appodize $m$ --- estimated from the 
timestream rms at the map making stage \citep[see][]{pryke09}.
For source identification we construct signal to noise maps as
\begin{equation}
\label{eq:s2n}
s=m'v^{-1/2}_{pix}.
\end{equation}
Though the filtering operation changes the noise amplitude, $v_{pix}$
provides information about the spatial distribution of the noise.
We rescale the amplitude of $s$ so that the $16^\textrm{th}$ percentile 
point of the pixel distribution is equal to $-1$.
To identify sources in $s$, we subtract a source template 
--- constructed from the back-transform of $W$ --- 
from the brightest pixels in $s$ and iterate down to a threshold value of 5.
This procedure results in a source free $s$-map with pixel values that are
close to Gaussian distributed in the range $\pm5\sigma$ and a catalog of 
$>5\sigma$ sources, where $\sigma\approx10$~mJy at both frequencies.

We generate a source catalog for both the real maps and for each of our 
point source injected signal plus noise simulated maps
and calculate the power spectra masking out the $>5\sigma$ sources.
The resulting distribution of the radio source injected simulations, 
as compared to $\Lambda$CDM alone, is plotted in Figure~\ref{fig:simband}.
Comparing the data against this new model we find that
the addition of residual radio sources marginally improves the
$\chi^2$ at both 100 and 150~GHz.
The bandpowers values, together with their covariance matrices and window 
functions, are available in numerical form at http://quad.uchicago.edu/quad.

\section{Implications for SZE Foregrounds}

\begin{figure*}
\begin{center}
\resizebox{\textwidth}{!}{\includegraphics{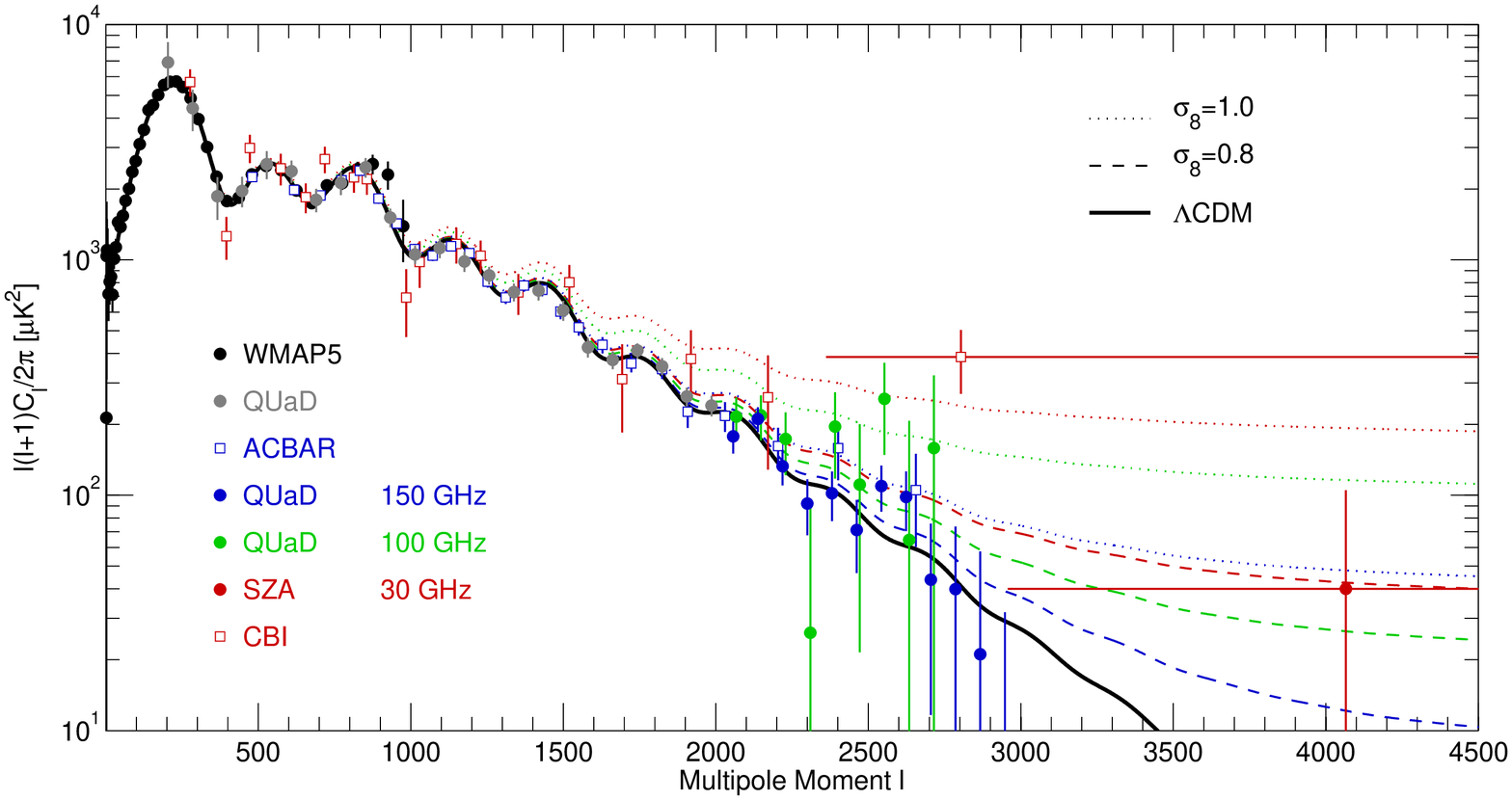}}
\end{center}
\caption{
The QUaD high-$\ell$ {\em TT} results for $2000<\ell<3000$
compared against recent results from 
ACBAR \citep{reichardt08}, CBI \citep{sievers09} and SZA \citep{sharp09} 
--- spanning the spectral range 30, 100 
and 1500~GHz ({\em red, green} and {\em blue} respectively) ---
plus WMAP \citep{hinshaw09} ({\em black}) and
QUaD \citep{brown09} for $\ell<2000$ ({\em grey}).  
For QUaD, SZA and CBI the estimated residual radio source contribution
has been subtracted.
Some points have been slightly offset in multipole for clarity.
The data are plotted against $\Lambda$CDM alone and $\Lambda$CDM 
plus the standard \cite{komatsu02} template assuming two values of 
$\sigma_{8}$ scaled to all three frequencies.
The QUaD data favors the $\sigma_{8}=0.8$ model with a best-fit scaling of 
$A_{SZ}=1.2\pm1.2$ (see text).
}
\label{fig:cmball}
\end{figure*}

Figure \ref{fig:cmball} shows the QUaD High-$\ell$ {\em TT} results along 
with measurements from other recent experiments.
There is broad agreement between the various datasets --- amongst each 
other and with $\Lambda$CDM --- except at the highest multipoles.
Here CBI \citep{sievers09} claims a significant excess,
which was ``confirmed'' by ACBAR \citep{reichardt08} at 1-$\sigma$,
whereas SZA \citep{sharp09} is consistent with zero power.
This intimation of excess power has been attributed to SZE 
signal and used to derive corresponding constraints on the 
amplitude of density perturbations $\sigma_8$.
Depending on the template used, the CBI (and to a lesser extent ACBAR) 
data imply values in the range $0.9<\sigma_{8}<1.0$, a departure from the 
conventional value of 0.8.

Following suit, we add the standard \cite{komatsu02} (KS) template 
--- scaled by a single parameter $A_{SZ}$ --- 
to $\Lambda$CDM plus the residual radio source contribution 
(Section~\ref{sec:psims}) and fit this model 
to the QUaD bandpowers.
The scale parameter can be related to 
a value of $\sigma_{8}$ as
\begin{equation}
\sigma_{8}^{KS}\approx0.8A_{SZ}^{1/7}.
\end{equation}

The 100 and 150 GHz data are considered both independently and 
simultaneously.
We use the bandpower covariance matrix from simulations
and add a beam plus absolute calibration systematic 
term calculated as
\begin{equation}
\textrm{M}'_{bb'}=\textrm{M}_{bb'}+
a^2\big(S_b\hat{C_b}\big)\big(S_{b'}\hat{C_{b'}}\big),
\end{equation} 
where $\hat{C}_b$ is the real bandpower value, $a$ is the absolute 
calibration uncertainty and $S_b$ is the beam uncertainty.
We calculate a $\chi^2$ for $A_{SZ}$ as
\begin{equation}
\chi^2=\big(\hat{C}_b - \langle C^{MC}_b \rangle - A_{SZ}C_b^{KS} \big)
~\textrm{M}'^{-1}_{bb'}~
\big(\hat{C}_{b'} - \langle C^{MC}_{b'} \rangle - A_{SZ}C_{b'}^{KS} \big),
\end{equation}
where $\hat{C}_b$, $C^{MC}_b$ and $C_b^{KS}$ are the observed, 
simulation and KS--expectation bandpowers respectively.
The sample variance component of the covariance matrix is scaled in 
proportion to the level of model signal akin to role of the more 
conventional offset-lognormal transformation.
The resulting likelihood is still very nearly Gaussian.

We note that the procedure adopted here neglects the non-Gaussianity of 
the SZE signal {\it and its fluctuation} and to correctly include this would 
require that we inject simulated SZE sky into our maps 
--- we have not done this.

We quote the maximum-likelihood value, 
with the 1-$\sigma$ uncertainties corresponding to
the likelihood falloff that encompasses 68\% of the total; 
95\% upper limits are evaluated 
by integrating the likelihood over postive values of $A_{SZ}$.
The results for 100 and 150~GHz are
$A_{SZ}=0.9\pm1.4$ and $1.0\pm1.5$ or $A_{SZ}<3.6$ and 3.8 
respectively with simultaneous fit values of
$A_{SZ}=1.2\pm1.2$ or $A_{SZ}<3.3$. 

These results are consistent with SZE power at the expected level for 
$\sigma_{8}=0.8$ but inconsistent with those of \cite{sievers09}, 
preferring lower values of $A_{SZ}$.
There is good agreement with the conclusions of \cite{sharp09}, who 
also make a strong argument that the CBI point source estimate is in fact
too small.
While the frequency dependence of the SZE makes measurements at 100 
and 150~GHz intrinsically less sensitive than those at 30~GHz,
they are also less prone to contamination by radio sources.

\section{Conclusions}

We have extended the range of the {\it TT}-bandpowers from QUaD to 
$\ell=3000$ using signal to noise weighting to downweight noisy regions of
the two-dimensional Fourier plane, an improved method for the removal
of ground contamination and higher accuracy beam modeling.
After masking point sources detected at high significance in our maps,
the spectra are consistent with the $\Lambda$CDM expectation alone.
We have estimated the residual radio source contribution using a 
physically motivated radio source model scaled to fit our bright 
source counts and find the contribution to be small at 100 GHz and 
negligible at 150GHz.

A small SZE contribution is expected at $\ell<3000$  for 
$\sigma_8=0.8$ (see Figure~\ref{fig:cmball}).
Fitting a standard SZE template spectrum to our data results in a best-fit 
amplitude consistent with the expectation.

\acknowledgements

We would like to thank Gianfranco De Zotti for providing us with his radio 
source counts at 100 and 150~GHz.
QUaD is funded by the National Science Foundation in the USA,
through grants AST-0096778, ANT-0338138, ANT-0338335 \&
ANT-0338238, by the UK Science and Technology Facilities Council (STFC)
and its predecessor the Particle Physics and Astronomy Research Council
(PPARC), and by the Science Foundation Ireland.
JRH acknowledges the
support of an NSF Graduate Research Fellowship, a Stanford
Graduate Fellowship and a NASA Postdoctoral Fellowship.
CP acknowledges partial support from
the Kavli Institute for Cosmological Physics through the grant NSF
PHY-0114422.  EYW acknowledges receipt of an NDSEG fellowship. YM
acknowledges support from a SUPA Prize studentship. PGC
acknowledges funding from the Funda\c{c}\~{a}o para a Ci\^{e}ncia e a Tecnologia.
MZ acknowledges support from a NASA Postdoctoral Fellowship.

\bibliographystyle{apj}
\bibliography{ms}

\begin{thebibliography}{21}
\expandafter\ifx\csname natexlab\endcsname\relax\def\natexlab#1{#1}\fi

\bibitem[{{Ade} {et~al.}(2008){Ade}, {Bock}, {Bowden}, {Brown}, {Cahill},
  {Carlstrom}, {Castro}, {Church}, {Culverhouse}, {Friedman}, {Ganga}, {Gear},
  {Hinderks}, {Kovac}, {Lange}, {Leitch}, {Melhuish}, {Murphy}, {Orlando},
  {Schwarz}, {O'Sullivan}, {Piccirillo}, {Pryke}, {Rajguru}, {Rusholme},
  {Taylor}, {Thompson}, {Wu}, \& {Zemcov}}]{ade07}
{Ade}, P. {et~al.} 2008, \apj, 674, 22, arXiv:0705.2359

\bibitem[{{Brown} {et~al.}(2009){Brown}, {Ade}, {Bock}, {Bowden}, {Cahill},
  {Castro}, {Church}, {Culverhouse}, {Friedman}, {Ganga}, {Gear}, {Gupta},
  {Hinderks}, {Kovac}, {Lange}, {Leitch}, {Melhuish}, {Memari}, {Murphy},
  {Orlando}, {O'Sullivan}, {Piccirillo}, {Pryke}, {Rajguru}, {Rusholme},
  {Schwarz}, {Taylor}, {Thompson}, {Turner}, {Wu}, \& {Zemcov}}]{brown09}
{Brown}, M.~L. {et~al.} 2009, ArXiv e-prints, 0906.1003

\bibitem[{{Castro} {et~al.}(2009){Castro}, {Ade}, {Bock}, {Bowden}, {Brown},
  {Cahill}, {Church}, {Culverhouse}, {Friedman}, {Ganga}, {Gear}, {Gupta},
  {Hinderks}, {Kovac}, {Lange}, {Leitch}, {Melhuish}, {Memari}, {Murphy},
  {Orlando}, {Pryke}, {Schwarz}, {O'Sullivan}, {Piccirillo}, {Rajguru},
  {Rusholme}, {Taylor}, {Thompson}, {Turner}, {Wu}, \& {Zemcov}}]{castro09}
{Castro}, P.~G. {et~al.} 2009, ArXiv e-prints, 0901.0810

\bibitem[{{Dawson} {et~al.}(2006){Dawson}, {Holzapfel}, {Carlstrom}, {Joy}, \&
  {LaRoque}}]{dawson06}
{Dawson}, K.~S., {Holzapfel}, W.~L., {Carlstrom}, J.~E., {Joy}, M., \&
  {LaRoque}, S.~J. 2006, \apj, 647, 13, arXiv:astro-ph/0602413

\bibitem[{{de Zotti} {et~al.}(2005){de Zotti}, {Ricci}, {Mesa}, {Silva},
  {Mazzotta}, {Toffolatti}, \& {Gonz{\'a}lez-Nuevo}}]{dezotti05}
{de Zotti}, G., {Ricci}, R., {Mesa}, D., {Silva}, L., {Mazzotta}, P.,
  {Toffolatti}, L., \& {Gonz{\'a}lez-Nuevo}, J. 2005, \aap, 431, 893,
  arXiv:astro-ph/0410709

\bibitem[{{Dunkley} {et~al.}(2009){Dunkley}, {Komatsu}, {Nolta}, {Spergel},
  {Larson}, {Hinshaw}, {Page}, {Bennett}, {Gold}, {Jarosik}, {Weiland},
  {Halpern}, {Hill}, {Kogut}, {Limon}, {Meyer}, {Tucker}, {Wollack}, \&
  {Wright}}]{dunkley08}
{Dunkley}, J. {et~al.} 2009, \apjs, 180, 306, 0803.0586

\bibitem[{{Gregory} {et~al.}(1994){Gregory}, {Vavasour}, {Scott}, \&
  {Condon}}]{pmn}
{Gregory}, P.~C., {Vavasour}, J.~D., {Scott}, W.~K., \& {Condon}, J.~J. 1994,
  \apjs, 90, 173

\bibitem[{{Hinderks} {et~al.}(2009){Hinderks}, {Ade}, {Bock}, {Bowden},
  {Brown}, {Cahill}, {Carlstrom}, {Castro}, {Church}, {Culverhouse},
  {Friedman}, {Ganga}, {Gear}, {Gupta}, {Harris}, {Haynes}, {Keating}, {Kovac},
  {Kirby}, {Lange}, {Leitch}, {Mallie}, {Melhuish}, {Memari}, {Murphy},
  {Orlando}, {Schwarz}, {Sullivan}, {Piccirillo}, {Pryke}, {Rajguru},
  {Rusholme}, {Taylor}, {Thompson}, {Tucker}, {Turner}, {Wu}, \&
  {Zemcov}}]{hinderks08}
{Hinderks}, J.~R. {et~al.} 2009, \apj, 692, 1221, 0805.1990

\bibitem[{{Hinshaw} {et~al.}(2009){Hinshaw}, {Weiland}, {Hill}, {Odegard},
  {Larson}, {Bennett}, {Dunkley}, {Gold}, {Greason}, {Jarosik}, {Komatsu},
  {Nolta}, {Page}, {Spergel}, {Wollack}, {Halpern}, {Kogut}, {Limon}, {Meyer},
  {Tucker}, \& {Wright}}]{hinshaw09}
{Hinshaw}, G. {et~al.} 2009, \apjs, 180, 225, 0803.0732

\bibitem[{{Hivon} {et~al.}(2002){Hivon}, {G{\'o}rski}, {Netterfield}, {Crill},
  {Prunet}, \& {Hansen}}]{hivon02}
{Hivon}, E., {G{\'o}rski}, K.~M., {Netterfield}, C.~B., {Crill}, B.~P.,
  {Prunet}, S., \& {Hansen}, F. 2002, \apj, 567, 2, astro-ph/0105302

\bibitem[{{Komatsu} \& {Seljak}(2002)}]{komatsu02}
{Komatsu}, E., \& {Seljak}, U. 2002, \mnras, 336, 1256, arXiv:astro-ph/0205468

\bibitem[{Lewis {et~al.}(2000)Lewis, Challinor, \& Lasenby}]{lewis99}
Lewis, A., Challinor, A., \& Lasenby, A. 2000, Astrophys. J., 538, 473,
  astro-ph/9911177

\bibitem[{{Mauch} {et~al.}(2003){Mauch}, {Murphy}, {Buttery}, {Curran},
  {Hunstead}, {Piestrzynski}, {Robertson}, \& {Sadler}}]{sumss}
{Mauch}, T., {Murphy}, T., {Buttery}, H.~J., {Curran}, J., {Hunstead}, R.~W.,
  {Piestrzynski}, B., {Robertson}, J.~G., \& {Sadler}, E.~M. 2003, \mnras, 342,
  1117, arXiv:astro-ph/0303188

\bibitem[{{O'Sullivan} {et~al.}(2008){O'Sullivan}, {Cahill}, {Murphy}, {Gear},
  {Harris}, {Ade}, {Church}, {Thompson}, {Pryke}, {Bock}, {Bowden}, {Brown},
  {Carlstrom}, {Castro}, {Culverhouse}, {Friedman}, {Ganga}, {Haynes},
  {Hinderks}, {Kovak}, {Lange}, {Leitch}, {Mallie}, {Melhuish}, {Orlando},
  {Piccirillo}, {Pisano}, {Rajguru}, {Rusholme}, {Schwarz}, {Taylor}, {Wu}, \&
  {Zemcov}}]{osullivan08}
{O'Sullivan}, C. {et~al.} 2008, Infrared Physics and Technology, 51, 277

\bibitem[{{Pryke} {et~al.}(2009){Pryke}, {Ade}, {Bock}, {Bowden}, {Brown},
  {Cahill}, {Castro}, {Church}, {Culverhouse}, {Friedman}, {Ganga}, {Gear},
  {Gupta}, {Hinderks}, {Kovac}, {Lange}, {Leitch}, {Melhuish}, {Memari},
  {Murphy}, {Orlando}, {Schwarz}, {Sullivan}, {Piccirillo}, {Rajguru},
  {Rusholme}, {Taylor}, {Thompson}, {Turner}, {Wu}, \& {Zemcov}}]{pryke09}
{Pryke}, C. {et~al.} 2009, \apj, 692, 1247, 0805.1944

\bibitem[{{Readhead} {et~al.}(2004){Readhead}, {Mason}, {Contaldi}, {Pearson},
  {Bond}, {Myers}, {Padin}, {Sievers}, {Cartwright}, {Shepherd}, {Pogosyan},
  {Prunet}, {Altamirano}, {Bustos}, {Bronfman}, {Casassus}, {Holzapfel}, {May},
  {Pen}, {Torres}, \& {Udomprasert}}]{readhead04b}
{Readhead}, A.~C.~S. {et~al.} 2004, \apj, 609, 498, arXiv:astro-ph/0402359

\bibitem[{{Reichardt} {et~al.}(2009){Reichardt}, {Ade}, {Bock}, {Bond},
  {Brevik}, {Contaldi}, {Daub}, {Dempsey}, {Goldstein}, {Holzapfel}, {Kuo},
  {Lange}, {Lueker}, {Newcomb}, {Peterson}, {Ruhl}, {Runyan}, \&
  {Staniszewski}}]{reichardt08}
{Reichardt}, C.~L. {et~al.} 2009, \apj, 694, 1200, 0801.1491

\bibitem[{{Sharp} {et~al.}(2009){Sharp}, {Marrone}, {Carlstrom}, {Culverhouse},
  {Greer}, {Hawkins}, {Hennessy}, {Joy}, {Lamb}, {Leitch}, {Loh}, {Miller},
  {Mroczkowski}, {Muchovej}, {Pryke}, \& {Woody}}]{sharp09}
{Sharp}, M.~K. {et~al.} 2009, ArXiv e-prints, 0901.4342

\bibitem[{{Sievers} {et~al.}(2009){Sievers}, {Mason}, {Weintraub}, {Achermann},
  {Altamirano}, {Bond}, {Bronfman}, {Bustos}, {Contaldi}, {Dickinson}, {Jones},
  {May}, {Myers}, {Oyarce}, {Padin}, {Pearson}, {Pospieszalski}, {Readhead},
  {Reeves}, {Shepherd}, {Taylor}, \& {Torres}}]{sievers09}
{Sievers}, J.~L. {et~al.} 2009, ArXiv e-prints, 0901.4540

\bibitem[{{Vikhlinin} {et~al.}(2009){Vikhlinin}, {Kravtsov}, {Burenin},
  {Ebeling}, {Forman}, {Hornstrup}, {Jones}, {Murray}, {Nagai}, {Quintana}, \&
  {Voevodkin}}]{vikhlinin08}
{Vikhlinin}, A. {et~al.} 2009, \apj, 692, 1060, 0812.2720

\bibitem[{{Wright} {et~al.}(2009){Wright}, {Chen}, {Odegard}, {Bennett},
  {Hill}, {Hinshaw}, {Jarosik}, {Komatsu}, {Nolta}, {Page}, {Spergel},
  {Weiland}, {Wollack}, {Dunkley}, {Gold}, {Halpern}, {Kogut}, {Larson},
  {Limon}, {Meyer}, \& {Tucker}}]{wright08}
{Wright}, E.~L. {et~al.} 2009, \apjs, 180, 283, 0803.0577

\end{thebibliography}

\end{document}